\documentclass[a4paper]{scrartcl}
\usepackage[margin=2.5cm]{geometry}
\usepackage{hyperref, booktabs, siunitx}
\DeclareSIUnit{\muas}{\mu\text{as}}
\DeclareSIUnit{\jy}{\text{Jy}}
\usepackage[nomarkers]{endfloat}
\usepackage{authblk}
\usepackage{amsmath, amssymb, mathrsfs, dsfont, nicefrac, mathtools, xcolor}
\usepackage{cleveref}
\usepackage{graphicx}
\usepackage{subcaption}
\usepackage[nomargin,inline,marginclue,draft]{fixme}
\usepackage[mode=buildmissing]{standalone}
\usepackage{csquotes}
\usepackage{xr}
\usepackage{tikz}
\usetikzlibrary{graphs, positioning, shapes.geometric, arrows.meta, shapes.multipart}
\usepackage{pgfplots}
\pgfplotsset{compat=1.14}

\usetikzlibrary{shapes}

\usepackage[backend=biber,style=nature]{biblatex}
\author[1,2]{Jakob Knollm\"uller}
\author[1,3]{Philipp Arras}
\author[2,3,4]{Torsten En{\ss}lin}
\affil[1]{Technical University of Munich, TUM School of Natural Sciences, Boltzmannstr.~2, 85748 Garching, Germany}
\affil[2]{Excellence Cluster ORIGINS, Boltzmannstr.~2, 85748 Garching, Germany}
\affil[3]{Max-Planck Institut f\"ur Astrophysik, Karl-Schwarzschild-Str.~1, 85748 Garching, Germany}
\affil[4]{Ludwig-Maximilians-Universit\"at M\"unchen (LMU), Geschwister-Scholl-Platz~1, 80539 M\"unchen, Germany}

\title{Resolving Horizon-Scale Dynamics of Sagittarius~A*}
\begin{document}
\maketitle

\begin{bfseries}

\noindent

\section*{Abstract}
\begin{textbf}
Sagittarius\,A* (Sgr\,A*), the supermassive black hole at the heart of our galaxy, provides unique opportunities to study black hole accretion, jet formation, and gravitational physics.
The rapid structural changes in Sgr\,A*'s emission pose a significant challenge for traditional imaging techniques.
We present dynamic reconstructions of Sgr\,A* using Event Horizon Telescope (EHT) data from April~6th and~7th, 2017, analyzed with a one-minute temporal resolution with the Resolve framework.
This Bayesian approach employs adaptive Gaussian Processes and Variational Inference for data-driven self-regularization.
Our results not only fully confirm the initial findings by the EHT Collaboration for a time-averaged source but also reveal intricate details about the temporal dynamics within the black hole environment.
We find an intriguing dynamic feature on April~6th that propagates in a clock-wise direction.
Geometric modelling with ray-tracing, although not fully conclusive, indicates compatibility with high-inclination configurations of about $\theta_o = 160^\circ$, as seen in other studies.
\end{textbf}

\end{bfseries}

Sagittarius A* (Sgr\,A*) is due to its proximity a critical observational target for probing the extreme physical conditions in the vicinity of supermassive black holes.
Early evidence of its black hole nature was derived from stellar kinematics near the galactic center \parencite{ghez1998high, gillessen2009monitoring}. 
With the Event Horizon Telescope (EHT), it was possible for the first time to directly image the black hole shadows of M87* \parencite{ehti,ehtii,ehtiii,ehtiv} and recently also Sgr\,A* \parencite{ehtsgra1,ehtsgra2,ehtsgra3,ehtsgra4,ehtsgra5}.
It leverages Very Long Baseline Interferometry (VLBI), which combines data from radio telescopes around the globe at frequencies of $230\,\text{GHz}$ to achieve such exceptional spatial resolutions.

Compared to relatively static targets, such as M87*, Sgr\,A* exhibits rapid variability on minute timescales, due to its small physical size.
This dynamic environment serves as a unique laboratory for exploring black hole phenomena, but it necessitates innovative techniques for dynamic reconstruction. 

In this paper, we use the Resolve framework  \parencite{resolve16, resolve18, resolve19} to address these challenges and provide a minute-by-minute dynamic reconstruction of Sgr\,A* using the EHT observations from April~6th and 7th, 2017.
Resolve is based on information field theory \parencite{ensslin09, ensslin18} and its numerical infrastructure \parencite{nifty1, nifty3, nifty5, niftycode}.
It employs adaptive Gaussian Processes and Variational Inference for Bayesian radio interferometry.

So far, the EHT Collaboration (EHTC) obtained the first, static, image of Sgr\,A*'s shadow by inflating the error budget to account for the time-variability and employing several imaging techniques \parencite{shepherd1997difmap,akiyama2017imaging,chael2018interferometric,broderick2020themis}.
Inflating error budgets for the reconstruction has two kinds of disadvantages. 
First, it degrades the quality of the image and second, the highly interesting dynamic source structure is lost.

Dynamic reconstruction, which tracks temporal changes in source structure, offers an alternative.
The main challenge is data sparsity per snapshot, as Earth's rotation cannot be relied on to fill in the measured Fourier-space.
The EHTC has explored dynamic reconstructions for a 100-minute window for both observations, employing geometric modeling \parencite{johnson2020universal} and the StarWarps algorithm \parencite{bouman2018reconstructing}.
Geometric models fit parametric structures to data, while non-parametric methods like StarWarps may unveil unexpected features.

StarWarps \parencite{bouman2018reconstructing} employs a probabilistic graphical model to trace the temporal evolution of a source.
In this approach, each frame is correlated with its successor based on a preset correlation strength.
The EHTC applied StarWarps to the reconstruction of Sgr\,A* and recovered intriguing structures.
Nevertheless, the results were deemed inconclusive due to the strong dependence on the correlation parameter.

The Resolve framework, which we are using in this work, uniquely adapts correlations across domains, such as between frames, based on the data.
This data-driven approach dynamically adjusts correlation strength, eliminating the need for fixed values.
Like all Bayesian methods, Resolve requires specifying measurement likelihoods, a model, and prior distributions.
Unlike newer methods using machine learning \parencite{terris2023image, medeiros2023image}, our approach is training-set independent, avoiding black-box pitfalls.

The likelihood quantifies the probability of obtaining a data set for a specific sky configuration. 
Various factors, from the instrumental setup to intermediary effects between the source and correlator, influence this likelihood.
In our case we employ an interferometric likelihood, where each antenna pair probes a specific spatial frequency of the observed source.
This corresponds to an individual point in Fourier space.
The EHT limits us to a sparse set of these points, resulting in highly incomplete sky coverage.

Radio waves from the Galactic center are affected by diffractive and refractive scattering in the interstellar medium.
The former is addressed by forward-modeling with a scattering kernel \parencite{johnson2018scattering}.
The latter, due to data limitations, is not explicitly accounted for, potentially affecting the small-scale features in our reconstructions.

The radio waves are also influenced by Earth's atmosphere in various ways, such as atmospheric delays or absorption.
We base our analysis on pre-calibrated data (HOPS pipeline \parencite{blackburn2019eht}), which mitigates many of these effects.
To account for remaining amplitude uncertainties and unreliable phases, we use self-calibration for the amplitudes and closure phases.
Closure phases are combinations of three visibility phases that are invariant under antenna-based phase gains, rendering them insensitive to calibration errors \parencite{closure19}.
The amplitude gains of the antennas can be determined with reasonable accuracy (see~\parencite[table\ 3]{ehtiii}).
For the self-calibration, we extend our model by incorporating amplitude calibration parameters for each visibility.
We then simultaneously solve for these parameters and the source structure itself to obtain self-consistent solutions.

Furthermore, simultaneous observations with the ALMA telescope yield light-curves for the total flux of Sgr\,A* \parencite{wielgus2022millimeter}.
For details on our likelihood implementation refer to \enquote{\nameref{sec:likelihood}} in the method section.

Moving on to the prior model, we want to infer the spatio-temporal evolution of the source structure, which is a three-dimensional function with two spatial and one temporal coordinate.
An infinite number of functions are compatible with any given data, necessitating the use of physical plausibility constraints.
To address this, we make three prior assumptions within the Bayesian framework.
The cornerstone of our approach is the assumption of spatial and temporal correlation.
We expect one location in the source to provide insight into neighboring points, with the degree of similarity diminishing with distance.
This correlation is formally expressed through Gaussian Processes, which are probability distributions over functions with characteristic correlation properties expressed in a kernel function.
Here our approach notably differs from the StarWarps method, as we incorporate correlation as a flexible part within our model and make it adaptable to the data instead of fixing it \textit{ab initio}.

Our second and third assumptions involve flux positivity and exponential changes on linear temporal and spatial scales.
Since radio emission traces density, negative emissions are un-physical and should be excluded.
For black holes, we expect sharp features like shadows and emission spanning orders of magnitude across the source.
Gaussian Processes do not inherently possess these properties, as they allow negative values and change linearly.
Using them to describe the logarithmic source brightness, we can address these limitations.
Overall, we use a log-normal Gaussian Process prior with an adaptive correlation structure to model the source.
For details on our prior assumptions refer to \enquote{\nameref{sec:prior}} in the method section.

To solve the Bayesian inference problem and thereby obtaining the reconstruction, we employ Variational Inference of the posterior distribution using Metric Gaussian Variational Inference (MGVI) \parencite{mgvi}.
This technique approximates the target distribution with a Gaussian distribution in a standardized space.
In contrast to traditional variational approaches, MGVI captures posterior correlations between all model parameters by exploiting the Fisher metric of the true posterior.
However, the true posterior distribution is likely multi-modal, a feature our uni-modal approximation cannot capture.
We therefore perform twelve independent reconstructions and average over the resulting samples to account for this additional uncertainty.
For details on MGVI refer to \enquote{\nameref{sec:mgvi}} in the method section.

The Resolve framework has been extensively validated in the context of VLBI and other radio interferometry applications. 
\textcite{arras2022variable} performs a systematic hyper-parameter study and found robust results throughout a wide range of prior parameter choices on a spatio-spectral-temporal reconstruction of M87*~EHT~2017 data.
Furthermore, the Resolve algorithm participated in the \emph{ngEHT Analysis Challenge~2} \parencite{ngehtAnalysisChallenge}.
There, a comparison of Resolve with several other approaches and various evaluation metrics is presented, utilizing simulated observations of Sgr\,A* as well as M87* GRMHD models for EHT2022 and envisioned ngEHT array configurations \parencite{doeleman2023reference}.
This exercise demonstrated Resolve's capability to robustly recover the dynamics of highly variable sources from EHT data.
Additionally, Resolve has proven effective in detecting the faint star S300 in close orbit around Sgr\,A* using the GRAVITY instrument \parencite{abuter2022deep}.
A more general validation of Resolve and its comparison to CLEAN \parencite{hogbom1974aperture} can be found in \textcite{arras2021comparison}.

\subsection*{Results}

\begin{figure}
    \centering
    \includegraphics[width =\textwidth]{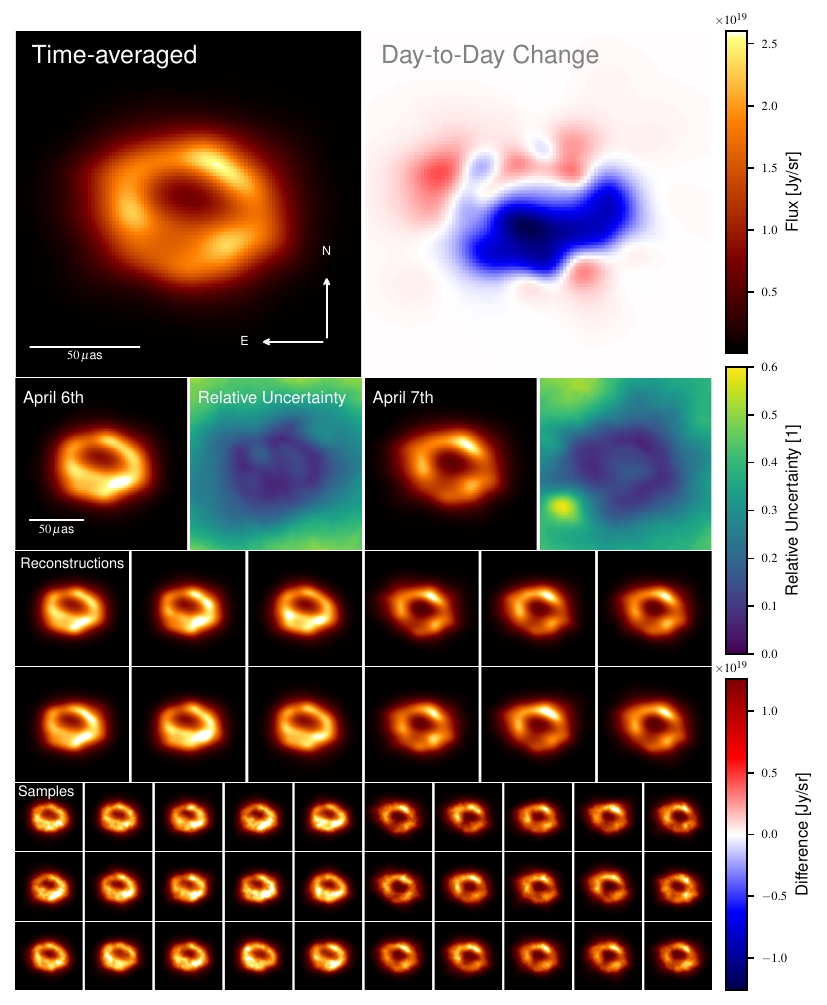}
    \caption[Time-averaged morphologies of Sgr\,A*]{Detailed visual analysis of the time-averaged morphologies of Sgr\,A*.
    The first row shows the temporal mean across both days, April~6th and~7th, derived from averaging all samples of all independent reconstructions together with the day-by-day change.
    The second row illustrates the mean reconstructions for each respective day together with their pixel-wise relative uncertainty.
    Rows three and four present a subset of the independent reconstructions, demonstrating their consistent morphology across different initializations. 
    Note that the left half shows April~6th, and the right April~7th.
    The final rows display randomly selected samples from the approximate posterior distributions, underscoring the variations in small-scale features due to remaining uncertainties. 
    }\label{fig:black-hole}
\end{figure}

\Cref{fig:black-hole} displays the time-averaged source morphology, revealing a consistent bright emission ring with a \SI{25 \pm 1}{\muas} diameter across all days and reconstructions. 
This feature is substantiated by pixel-wise uncertainties below \SI{10}{\percent}. Slight deviations from circular symmetry are evident, indicated by an ellipticity parameter $\tau=0.34 \pm 0.05$. 
\Cref{tab:simple_table} contains a number of source parameters extracted with VIDA \parencite{tiede2022vida}.
For details refer to \enquote{\nameref{sec:VIDA}} in the method section.
These observations generally align with EHTC findings \parencite{ehtsgra3}.

Distinct morphologies are observed between April~6th and 7th; the former features a bright spot in the southwest, the latter has three bright regions.
Our April~7th reconstruction aligns with EHTC results \parencite{ehtsgra1}, differing only in faint structures surrounding the source.

As no summary image for April~6th is available from EHTC publications, we compare our reconstruction to the method averages within the EHTC.\@
For April~6th, we compare our reconstruction with EHTC's on-sky HOPS column as seen in \parencite[fig.\ 17]{ehtsgra3}, where we find morphological agreement with the general structure of THEMIS.

These inter-day variations are consistent with the source's variability. 
Overall, our work corroborates EHTC findings and supports the presence of a supermassive black hole at the Galactic center.

The third and fourth rows in \cref{fig:black-hole} display time-averaged individual reconstructions for April~6th and 7th, illustrating the effect of our posterior's multi-modality.
Despite minor flux variations, the morphology remains consistent across reconstructions. 
The final rows feature samples from the approximate posteriors. 
These samples display small-scale features not data-constrained but arising from priors. 
Averaging these smoothes out such features and contributes to the observed uncertainty.

\begin{figure}
    \centering
    \includegraphics[width =0.9\textwidth]{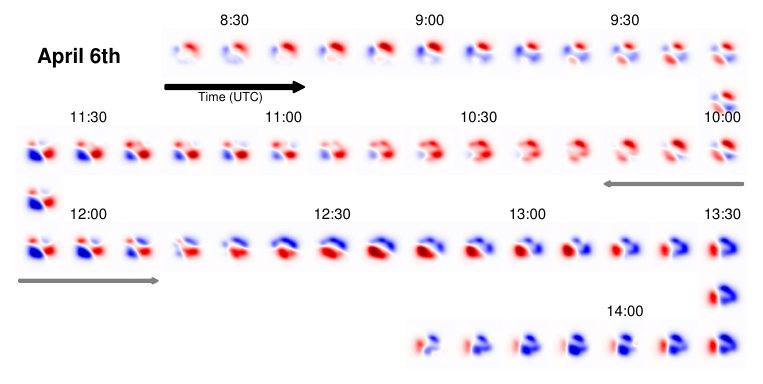}
    \includegraphics[width =0.9\textwidth]{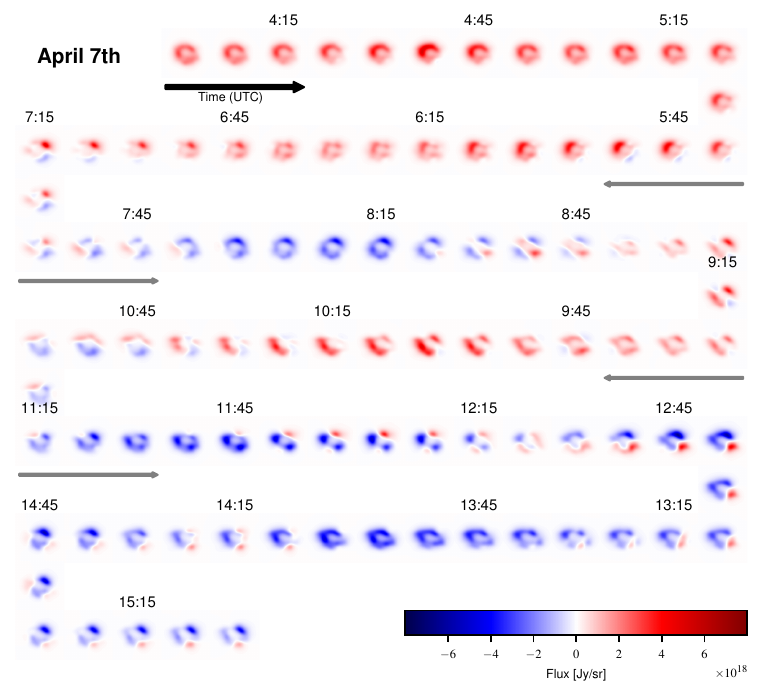}
    \caption[Spatio-temporal Evolution of Sgr\,A*]{The figure portrays the spatio-temporal evolution of Sgr\,A* across the two observation dates, April~6th and~7th, with the corresponding day-median subtracted.
      Each snapshot, spaced \SI{7}{\minute} apart offers a detailed view of the dynamic changes over time.
      The temporal progression follows the arrow indicators, alternating between right and left directions from the top left corner.}\label{fig:movie}
\end{figure}

\Cref{fig:movie} details the source's temporal evolution during April~6th and~7th. We subtracted the temporal median image of the corresponding day to emphasize intra-day changes.
Note that the number of available antennas changes throughout the day \parencite{ehtsgra2}. 
The best coverage falls between about 11:00 UTC and 14:30 UTC, where we expect the hightest fidelity of the reconstruction and its dynamics.
Times outside this window might be predominantly informed by the average source structures, as it might not be possible to discern any meaningful variable structure.
This can be directly seen on April~7th, as the a flux initially does not show much localized dynamics.
With increased coverage, we observe more localized transient features, but we cannot report coherent motion.

\begin{figure}
    \centering
    \includegraphics[width =\textwidth]{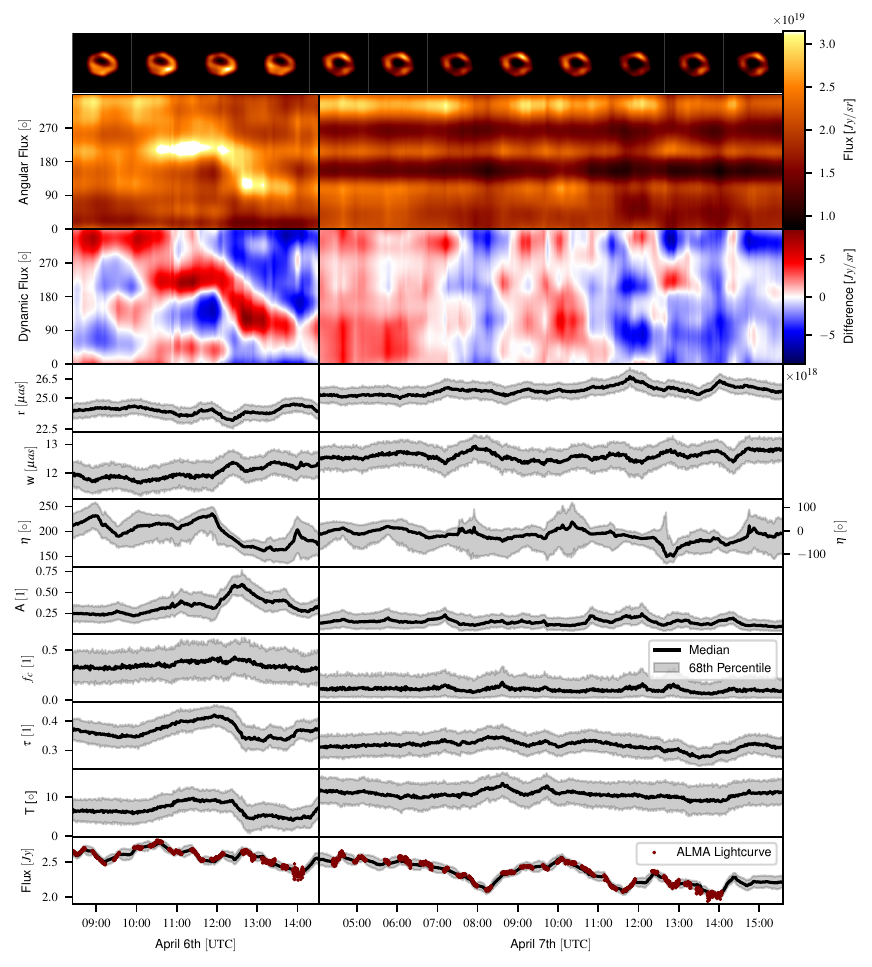}
    \caption[Temporal Evolution of Sgr\,A* Parameters and Ring]{This figure demonstrates the temporal evolution of Sgr\,A*'s parameters and ring on April~6th and~7th.
    The first row presents frames of the reconstruction, captured at equidistant intervals, to depict the source state over time. 
    The second row visualizes the mean angular flux of the unrolled ring, moving from north in the east direction. 
    The third row delineates the dynamic flux, calculated by subtracting the temporal median of the respective day. 
    Rows four to ten portray the sample median and the 68th~percentile range of the cosine ring with a floor model parameters, as determined by the VIDA (for details refer to \enquote{\nameref{sec:VIDA}} in the method section).
    The final row shows the total flux of the source alongside corresponding ALMA observations~\parencite{wielgus2022millimeter}.}\label{fig:temporal}
\end{figure}

This is in contrast to April~6th, where we do detect potentially coherent motion near Sgr\,A*'s horizon. 
Starting at 10:30 UTC, a bright spot emerges in the southwest and appears to move clockwise until about 13:30 before dissipating.
In \cref{fig:temporal}, we track angular flux evolution, black hole parameter changes, and total flux.
On April~6th, we observe a sharp localized increase in brightness at \SI{220}{\degree} and 10:30 UTC, followed by a clockwise shift from 12:00 UTC onward.

In \cref{fig:temporal}, April~7th exhibits stable angular flux patterns, including a prominent bright spot at \SI{336}{\degree}.
While similar features are present on April~6th, they are less distinct and overlaid with dynamic elements.
Most source parameters are generally stable; however, the ring asymmetry angle, $\eta$, differs between days.

The dynamic feature on April~6th affects several parameters, including asymmetry $A$, asymmetry angle $\eta$, ellipticity $\tau$, and ellipticity angle $T$.
The low variance in $\eta$ across reconstructions points to a consistent shift in ring brightness throughout the reconstructions.
The last row of \cref{fig:temporal} shows the total flux of our reconstruction together with the concurrent ALMA light-curve, which we track perfectly.

Several factors besides source evolution could contribute to the dynamic feature observed on April~6th.
Instrumental effects, such as fluctuating numbers of operating antennas, are one possibility.
Nevertheless, the feature appears precisely during optimal antenna coverage, and from 11:00\ UTC, the antenna count remains stable.
Another possibility is limitations inherent to our reconstruction method.
However, our prior model is only biased towards smooth structures, not continuous motion along circular paths.
Given the absence of similar dynamics on April~7th and the proven robustness of our approach, we attribute this feature more likely to the data itself, although unaccounted-for instrumental factors cannot be ruled out.
While our reconstruction does not account for a static scattering screen, its static nature makes it an unlikely source of the observed dynamic feature.

For the rest of our discussion, we assume this feature to be of physical origin and we explore its potential properties.
Observed dynamics around supermassive black holes like Sgr\,A* are complex, with multiple possible origins. 
One could be a localized emitter, such as a hot-spot, orbiting the black hole. 
Alternatively, large-scale structural evolution could induce apparent motion. 
To further examine this dynamic feature, we employ a temporally evolving hot-spot model. 
We isolate the feature by subtracting the temporal median and removing negative values. 
The resulting feature is then compared with a ray-traced model using the BAM package \parencite{palumbo2022bayesian}.

Our model, adapted from \textcite{palumbo2022bayesian}, is a simplistic, time-evolving hot-spot.
It uses a flat disk with Gaussian and van-Mises distributions for the radial and angular profiles, respectively.
The angular evolution is a linear function in time, augmented with parameters for source shift and Gaussian blur to consider observational artifacts.
A Gaussian likelihood with a conservative pixel-wise standard deviation accounts for extraction and modelling inaccuracies.
For parameter estimation, we use Dynamic Nested Sampling \parencite{skilling2006nested, higson2019dynamic}.
For details refer to \enquote{\nameref{sec:hotspot}} in the method section.

\begin{figure}
    \centering
    \includegraphics[width =\textwidth]{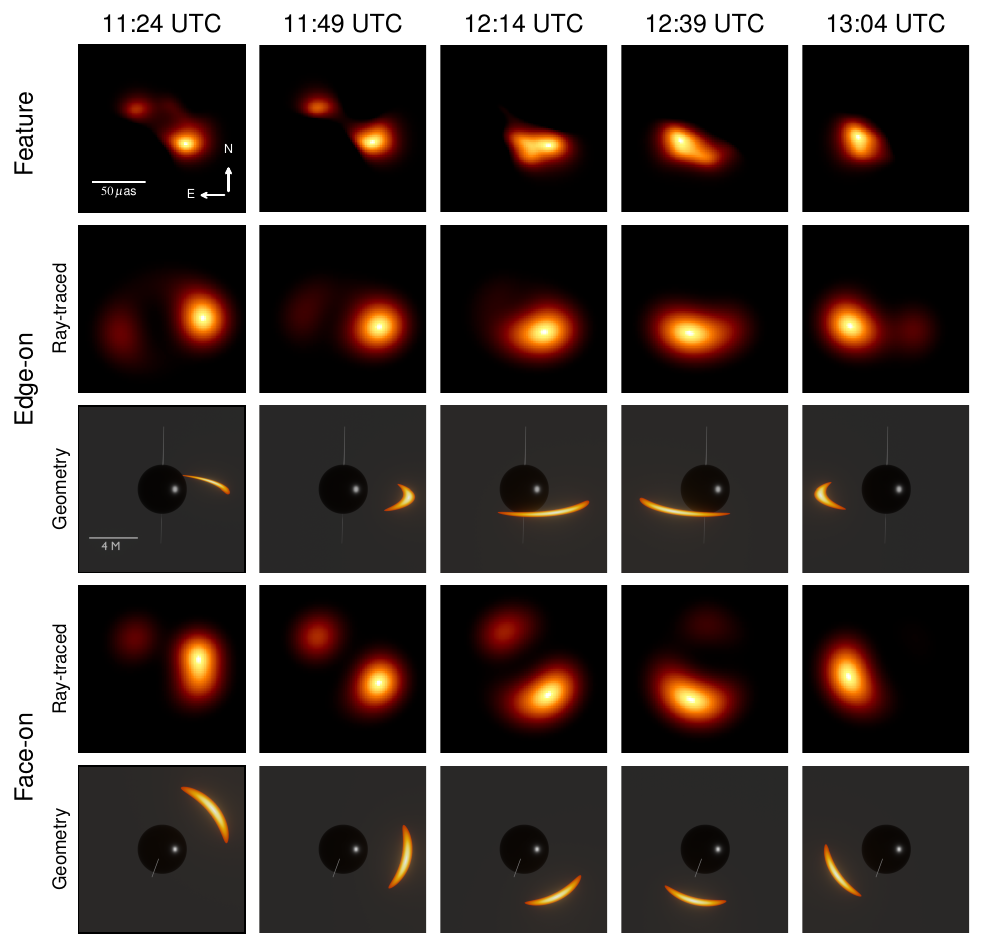} 
    \caption[Dynamic Feature and Best-Fit Models for Edge-On and Face-On.]{
        The figure presents the dynamic feature extracted from our reconstructions, alongside best-fit models for both edge-on and face-on orientations at different times. 
        The top row displays frames from our reconstruction after median subtraction and zero-clipping. 
        The second and third rows are dedicated to the edge-on orientation; the former showcases ray-traced and blurred models, while the latter illustrates the corresponding un-lensed 3D geometry. 
        The fourth and fifth rows offer analogous representations but for the face-on orientation.
        }
        \label{fig:hotspot}
\end{figure}

\begin{figure}
    \centering
    \includegraphics[width =\textwidth]{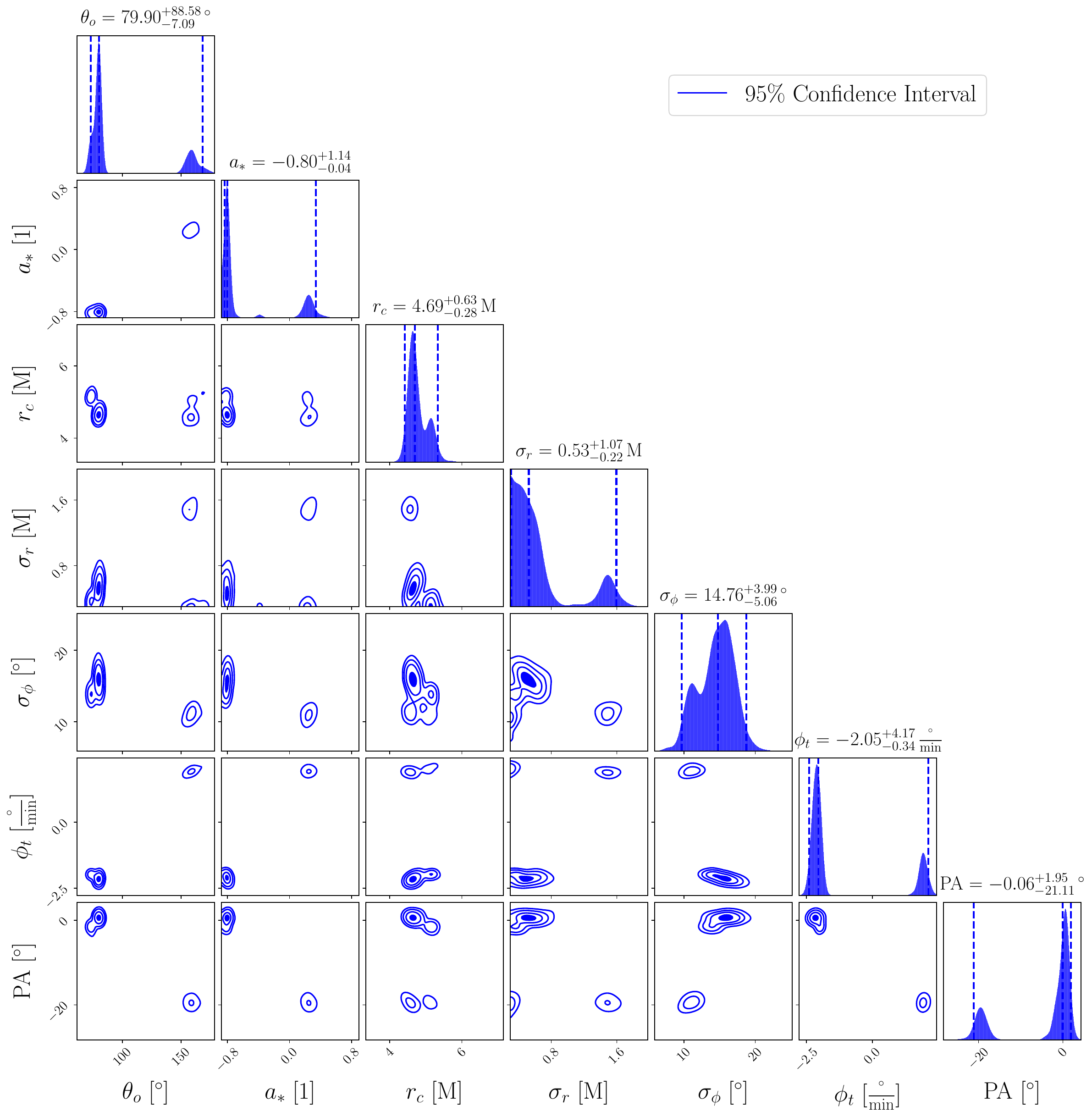}
    \caption[Corner plot for the geometric analysis.]{
        Corner plot of selected parameters from the nested sampling posterior.
        The edge-on mode corresponds to the larger and the face-on mode to the smaller peak in the inclination $\theta_0$.
        Uncertainties are provided as $95\%$ percentiles. 
        Their details are summarized in \cref{tab:geometric_parameters}.
    }
    \label{fig:cornerplot}

\end{figure}

Our model posterior reveals two distinct modes, corresponding to edge-on and face-on configurations of the observed feature.
The edge-on mode has an inclination $\theta_o = 79.6^{+1.2 \:\circ}_{-6.9}$ and position angle $\mathrm{PA}={0.4}^{+1.6\:\circ}_{-3.0}$; the face-on mode exhibits $\theta_o = 159.5^{+14.3 \:\circ}_{-8.1}$ and $\mathrm{PA}={-19.5}^{+2.9\:\circ}_{-3.2}$.
Visualizations and parameter details are featured in \cref{fig:hotspot}, \cref{fig:cornerplot}, and \cref{tab:geometric_parameters}.

While the posterior assigns $76.5\%$ probability mass to the edge-on configuration and $23.5\%$ to the face-on, both scenarios remain viable.
The edge-on configuration provides a more accurate representation of the primary bright spot's motion, whereas the face-on configuration encapsulates additional secondary features in the north-eastern quadrant.
It should be noted that the model's relative weighting is contingent upon the chosen likelihood function, limiting its interpretive value.

Both the extracted feature and the model configurations manifest a clockwise motion in projected sky coordinates.
However, these configurations imply divergent three-dimensional orbital dynamics.
In the edge-on scenario, we face the top side of the emission feature as it passes in front of the black hole. 
In the face-on scenario, the observer faces the underside of the feature as it traverses slightly behind the black hole. 

The inferred characteristic radius for both modes $r_c \approx 4.7\,\mathrm{M}$ falls below the innermost stable circular orbit (ISCO) $r_{\mathrm{ISCO}} = 6\,\mathrm{M}$ for Schwarzschild black holes, 
where $\mathrm{M}$ corresponds to its mass and natural units are used (speed of light $c= 1$ and gravitational constant $G=1$).
This would indicate a pro-grade spin of at least $|a|=0.37$, assuming no significant emission below ISCO. 
For both modes we do indeed find an alignment of the spin direction with the orbit of the feature.
The uncertainties on the spins would be consistent with this interpretation.

The absolute angular velocity for both modes is approximately $2\frac{ ^\circ}{\mathrm{min}}$, suggesting an orbital period of three hours. 
However, such a period is inconsistent with orbital velocities of $0.056\,c$ this close to Sgr\,A*. 
Expected speeds near Sgr\,A*'s (Schwarzschild-)ISCO are $0.41\,c$, far exceeding our observation. 
Our angular velocities would be consistent with a physical velocity of $0.23\,c$  expected for localized emission at a radius of $r_c \approx 19.2\, \mathrm{M}$. 
This radius, however, lies outside the geometrically determined radius range. 

Therefore, if we assume the feature to originate from localized emission orbiting the black hole, either our recovered radius or the angular velocity must be unreliable.
This could easily be due to our simplistic model or the extraction procedure and will need further investigation in the future.
However, given the absence of a clear flare signal during the relevant time period \parencite{wielgus2022orbital}, we are inclined to disregard this orbiting hot-spot scenario.

It is also possible that we do not observe a localized emission feature and instead the observed dynamics originate from more complex source behavior.
Macroscopic changes within the black hole system could lead to apparent continuous motion at speeds not necessarily related to orbital velocities. 
The feature also does not necessarily need to originate within the plane of the disk, it could be associated to the jet or the outside of the disk. 
In this case the dynamics are significantly more complex and we cannot use simple geometric relations to derive conclusions, and a more extensive study based on simulations is required.

Interestingly, the face-on mode aligns well with GRAVITY's infra-red polarimetric studies \parencite{abuter2018detection,baubock2020modeling} and ALMA observations \parencite{wielgus2022orbital}. 
Both suggest high-inclination angles around $\theta_o \approx 160 ^\circ$ and clockwise sky motion.
This face-on configuration also agrees with the findings of the EHTC by comparing the data to a model library of GRMHD simulations \parencite{ehtsgra5}.

In this study, we employed the Resolve framework to perform dynamic imaging of Sgr\,A* at a temporal resolution of one minute. 
This Bayesian approach allowed us to capture the complex spatio-temporal interplay within the horizon scales of the black hole.
Our findings fully confirm the initial results by the EHTC. 
We clearly recover a bright emission ring with central depression for both days of observation.
Extending the reconstruction into the temporal domain, we identify a dynamic feature on April~6th, 2017.
This feature underlines the results of previous studies suggesting high-inclination configurations and an on-sky clockwise motion with completely orthogonal methodology. 

Besides these findings, we presented the first spatially resolved dynamic features in the immediate vicinity of Sgr\,A*, the central black hole of our Galaxy. 
We no longer have to picture it as a static entity as we provided the first glimpse into its ever changing nature.

\section*{Methods}\label{sec:methods}

\begin{figure}
    \begin{tikzpicture}[>=stealth, node distance=2.5cm,
                    every node/.style={circle, draw, minimum size=1.5cm, font=\sffamily},
                    data node/.style={ellipse, draw, minimum width=2.5cm, minimum height=1.5cm, fill=gray!20},
                    calc node/.style={rectangle, rounded corners, fill=blue!20},
                    rel node/.style={diamond, fill=red!20},
                    elliptical node/.style={ellipse, draw, minimum width=2.5cm, minimum height=1.5cm, fill=green!20}]
    \node[data node] (d_amp) at (0,0) {$d_{\rho}$ \\ $\sim \mathcal{N}(d_{\rho} | Gg + \rho, N_{\rho})$};  
    \node[data node] (d_ph) at (5.5,0) {$d_{\varphi}$ \\ $\sim \mathcal{N}(d_{\varphi} | \varphi, N_{\varphi})$}; 
    \node[data node] (d_ALMA) at (9,4.) {$d_{\mathrm{lc}}$ \\ $\sim \mathcal{N}(d_{\mathrm{lc}} | s_t, N_{\mathrm{lc}})$}; 
    \node[calc node] (R_amp) at (2,2.5) {$\rho = \mathrm{ln}||\nu||$};  
    \node[calc node] (R_ph) at (5.5,2.5) {$\varphi = e^{iM \mathrm{atan}(\nu)}$};  
    \node[calc node] (R_ALMA) at (9,6.25) {$s_t = \int dx \:s$};  
    \node[calc node] (v) at (4.5,5) {$\nu = R B s$};  
    \node[calc node] (s) at (4.5,7.5) {$s = b \frac{e^{\tau}}{\frac{1}{T}\int dt\:dx\: e^\tau}$};  
    \node[calc node] (equation1) at (-2,2.5) {$g = \sigma_g \xi_g + \mu_g$};  
    \node[elliptical node] (equation2) at (-2,6) {$\xi_g \sim \mathcal{N}(\xi_g | 0, \mathds{1})$};  
    \node[calc node] (equation_a) at (0,10) {$b = e^{\sigma_b \xi_b + \mu_b}$};  
    \node[elliptical node] (equation3) at (0,12.5) {$\xi_b \sim \mathcal{N}(\xi_b | 0, 1)$};  
    \node[calc node] (equation_tau) at (4.5,10) {$\tau = A \xi_\tau$};  
    \node[elliptical node] (equation4) at (4.5,12.5) {$\xi_\tau \sim \mathcal{N}(\xi_\tau | 0, \mathds{1})$};  
    \node[calc node] (equation_A) at (9,10) {$A(\xi_A) = A_t \otimes A_x$, see \cite{arras2022variable}};  
    \node[elliptical node] (equation5) at (9,12.5) {$\xi_A \sim \mathcal{N}(\xi_A | 0, \mathds{1})$};  
    
    \draw[->, thick] (R_amp) -- (d_amp);
    \draw[->, thick] (R_ph) -- (d_ph);
    \draw[->, thick] (R_ALMA) -- (d_ALMA);
    \draw[->, thick] (s) -- (v);
    \draw[->, thick] (v) -- (R_amp);
    \draw[->, thick] (v) -- (R_ph);
    \draw[->, thick] (s) -- (R_ALMA);
    \draw[->, thick] (equation1) -- (d_amp);  
    \draw[->, thick] (equation2) -- (equation1);
    \draw[->, thick] (equation_a) -- (s);
    \draw[->, thick] (equation3) -- (equation_a);
    \draw[->, thick] (equation_tau) -- (s);
    \draw[->, thick] (equation4) -- (equation_tau);
    \draw[->, thick] (equation_A) -- (equation_tau);
    \draw[->, thick] (equation5) -- (equation_A);
\end{tikzpicture}
    \caption[The graphical structure of the probabilistic model.]{
        Visual representation of the probabilistic model's graphical structure.
        Elliptic nodes signify stochastic quantities. 
        Observed quantities are indicated by gray, while the inferred ones are in green. 
        Blue square nodes, coupled with arrows pointing towards the generative direction, depict deterministic relationships established by the forward model. 
        For further understanding of the mathematical symbols, refer to the methods section. 
        In-depth details of the amplitude model are available in the methods section of \textcite{arras2022variable}.}\label{fig:graph}
\end{figure}
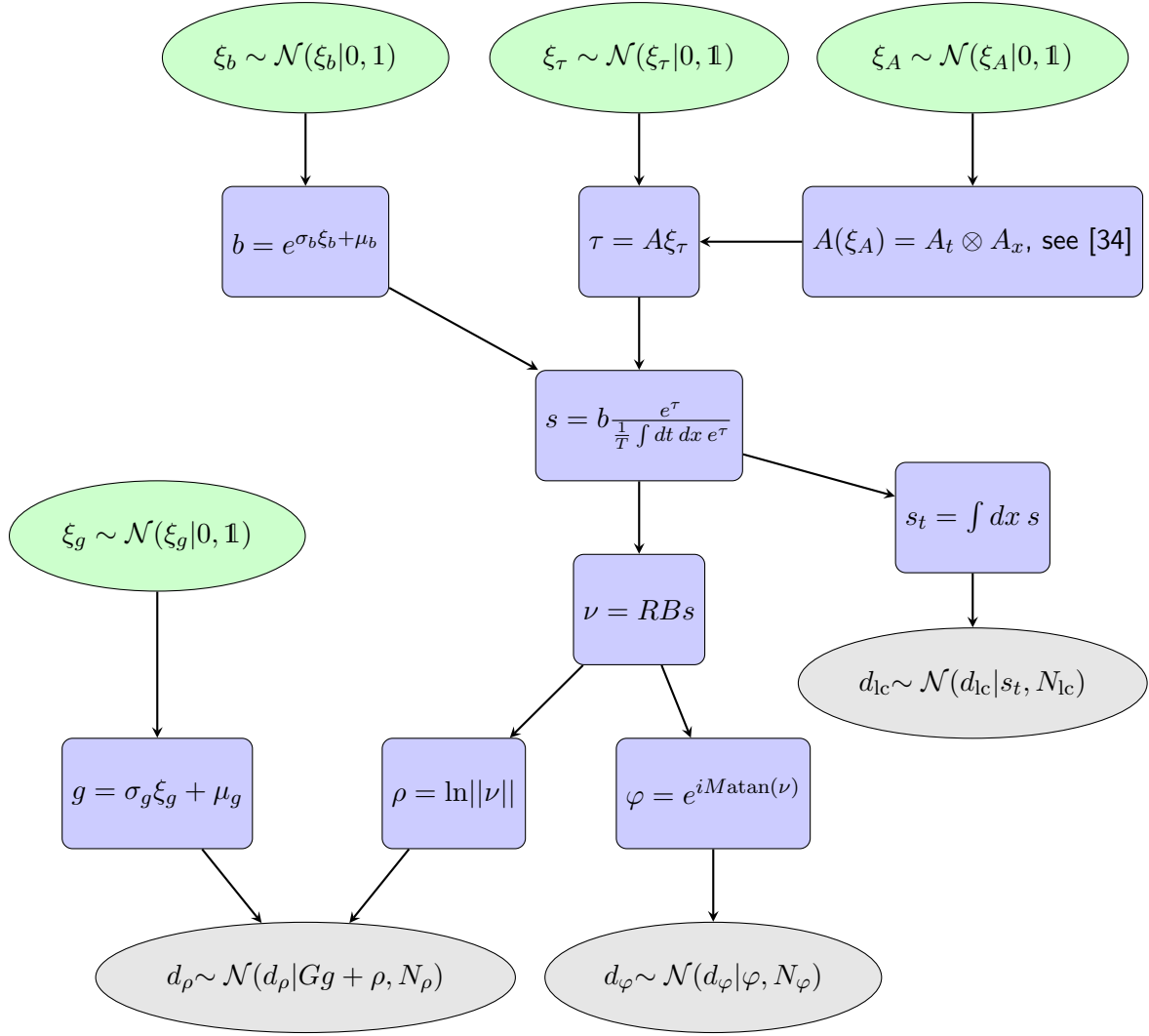

\paragraph{Bayesian imaging}
In Resolve we treat the imaging task as a Bayesian inference problem. 
We are interested in the probability distribution of image parameters $\theta$ that are compatible with the observed data $d$ as well as plausible under prior considerations, the so-called posterior $\mathcal{P}(\theta | d)$.
For this we have to combine the likelihood $\mathcal{P}(d | \theta)$ of the data given model parameters with the prior distribution $\mathcal{P}(\theta)$ according to Bayes' theorem:
\begin{align}
    \mathcal{P}(\theta|d) = \frac{\mathcal{P}(d | \theta) \mathcal{P}(\theta)}{\mathcal{P}(d)}
\end{align}
Here the computation of the evidence $\mathcal{P}(d)$ is usually not tractable for the kinds of nonlinear models we employ for the imaging.
Hence, we resort to approximations of the posterior distribution.
In dealing with our models, we encounter a large-scale inference problem that involves probability distributions over tens of millions of parameters. 
Traditional methods like Markov Chain Monte Carlo (MCMC) struggle to manage even when the number of parameters is much lower, in the range of tens to hundreds.
For a comprehensive overview and visual guide to the interplay between the prior and likelihood, refer to \cref{fig:graph}.

\paragraph{Likelihood}\label{sec:likelihood}
To carry out our reconstruction, we rely on data collected by the Event Horizon Telescope (EHT) on April~6th and~7th, 2017. 
This dataset informs us about the time-dependent behavior of the source, Sgr\,A*. 
Additional light-curve data from the ALMA array offer more information on the total source flux.

In statistical terms, a likelihood represents the probability of observing a specific dataset given a signal of interest, denoted by $s$. 
For our study, $s$ encapsulates a time-dependent description of the sky, detailed in the subsequent section. 
Therefore, it's crucial to have a model that faithfully follows the path of the emitted radiation from its source to its point of capture by the observing instrument.

Upon emission, the radiation initially encounters diffractive scattering caused by rapid variations in electron densities along the line of sight.
This scattering forms the first obstacle in the radiation's path, altering its characteristics as it moves towards the instrument.
This effect can be represented mathematically through a convolution operation with a scattering kernel \parencite{johnson2018scattering}. 
In the forward model, this convolution operation is denoted as $B$, meaning the instrument sees not the original source $s$, but a scattered version $s' = B s$. 
We use the assumed convolution kernel from \textcite{johnson2018scattering}, as implemented in eht-imaging \parencite{chael2018interferometric}. 
In addition to diffractive scattering, the radiation emission also undergoes refractive scattering, a process which could introduce small-scale features. 
However, given that the scope of our work is centered on the reconstruction of on-sky images, we will not incorporate this effect explicitly, in alignment with several EHTC reconstructions.\@

Other factors, like atmospheric conditions and antenna specifics, also affect the observed data. 
However, many of these issues can be reduced in the initial data calibration stage. 
For our study, we use data that has been calibrated using the HOPS pipeline \parencite{blackburn2019eht}.

The so-called sky visibility $\nu_{\mathrm{AB}}$ of an antenna pair $\mathrm{A}$ and $\mathrm{B}$ of the EHT at a given time probes a single Fourier coefficient of the sky flux distribution.
The corresponding measurement equation is expressed as:
\begin{align}
\nu_{\mathrm{AB}} = \int e^{-2\pi i\left(u_{\mathrm{AB}}x+v_{\mathrm{AB}}y\right)} s'(x,y) \,dx\, dy 
\end{align}
These visibilities, $\nu_{\mathrm{AB}}$, are complex numbers located at Fourier coordinates $u_{\mathrm{AB}}$ and $v_{\mathrm{AB}}$. 
These coordinates are determined by the distance between antennas A and B, scaled by the observational wavelength.
Despite rigorous calibration procedures, instrumental imperfections persist. 
These can be accounted for by adding another set of complex numbers that can modify the model visibilities.

In our study, we focus only on antenna-based effects and do not consider baseline-based ones. 
Mathematically this corresponds to a multiplication with complex numbers $\eta_\mathrm{A}$ and $\eta_\mathrm{B}$ for each involved antenna.
Furthermore we assume additive Gaussian noise $n_{\mathrm{AB}}$, originating from the thermal properties of the receiver.
Incorporating these considerations, the data equation becomes:
\begin{align}
d_\nu^{(\mathrm{AB})} = \eta_\mathrm{A} \eta_\mathrm{B}^* \nu_{\mathrm{AB}} + n_{\mathrm{AB}}
\end{align}
This leads to a complex-valued Gaussian likelihood for all measured visibilities, represented by $d_\nu$. 
The mathematical form of this likelihood is:
\begin{align}
    d_\nu^{(\mathrm{AB})} \sim \mathcal{N}(d_\nu^{(\mathrm{AB})}|\eta_\mathrm{A} \eta_\mathrm{B}^* \nu_{\mathrm{AB}}, N_\nu^{(\mathrm{AB})}) \quad ,
\end{align}
We use a diagonal noise covariance matrix $N_\nu^{(\mathrm{AB})}$ and take the inverse of the observational data's weights as our measure of uncertainty.

Here we will not work with the visibilities directly, as the phase calibration for VLBI networks in general and at these high frequencies in particular is excessively difficult, whereas the amplitudes can be constrained to a reasonable degree.
Instead we express the complex visibilities in terms of log-amplitudes $\rho$ and phases $\phi$, $\nu_{\mathrm{AB}} = e^{\rho_{\mathrm{AB}}+i\phi_{\mathrm{AB}}}$.
Given sufficiently high signal-to-noise ratios, we can linearize the likelihood around the observed value.
This yields a Gaussian distribution for the log-amplitudes:
\begin{align}
    \mathcal{P}(d_\rho^{(\mathrm{AB})}| \rho_{\mathrm{AB}}, g_\mathrm{A},g_\mathrm{B}) = \mathcal{N}(d_\rho^{(\mathrm{AB})}|\rho_{\mathrm{AB}} + g_\mathrm{A} + g_\mathrm{B}, N^{(\mathrm{AB})}_\rho) \quad \quad \text{with}\quad N^{(\mathrm{AB})}_\rho = \frac{N^{(\mathrm{AB})}_\nu}{|d_\nu^{(\mathrm{AB})}|^2}
\end{align}
The distribution can also be expressed compactly in vector notation:
\begin{align}
    \mathcal{P}(d_\rho| \rho, g) = \mathcal{N}(d_\rho|\rho + Gg, N_\rho)
\end{align}
In this context, $G$ is a sparse matrix that specifies the pair of antennas involved in constructing a particular baseline.
For the reconstruction we will parametrize the unknown logarithmic antenna gain amplitudes $g$ and use this likelihood as a part of the overall likelihood.

For phase information, we rely on closure phases, which are invariant to antenna-based phase gains.
In the early stages of the reconstruction, we also utilize closure log-amplitudes.
However, as the reconstruction progresses, we transition to self-calibrating the log-amplitudes.

To calculate likelihoods for closure log-amplitudes and closure phases, we begin by forming closure quantities from the measured visibilities.
Closure amplitudes are formed by combining the logarithmic absolute values of four visibilities using the closure matrix $L$, while closure phases are obtained by combining a triplet of complex phases of visibilities using the closure matrix $M$. 
The matrices consist of entries with $\pm 1$ for visibilities involved in the closure, and zero elsewhere.
Closure phases, denoted as $\phi_{\mathrm{cl}} = M \phi$, are calculated by applying the closure matrix $M$ to visibility phases.
To avoid complications due to phase wraps, we revert closure phases back to complex numbers $\varphi = e^{i M \phi}$.
This allows for the likelihood to be linearized around observed values for sufficiently high signal-to-noise ratios.

The resulting likelihood is then again a Gaussian distribution of the form:

\begin{align}
    \mathcal{P}(d_\varphi\vert\varphi) = \mathcal{N}(d_\varphi|\varphi, N_\varphi) \quad \quad \text{with}\quad  N_\varphi = \widehat{e^{-i M d_\phi}}M^T \frac{N_\nu}{|d_\nu|^2} M \widehat{e^{i M d_\phi}}
\end{align}

In the covariance, hats are used to denote diagonal matrices,
with the diagonal specified by the vector below the hat.
This is our second likelihood function used for reconstruction.
It is important to note that we do not compute the full, redundant set of closure phases. 
Instead, we opt for a non-redundant set derived from the baselines with the highest signal-to-noise ratio.

The likelihood for closure amplitudes is similarly a Gaussian:
\begin{align}
    \mathcal{P}(d_\varrho\vert\varrho) = \mathcal{N}(d_\varrho|\varrho, N_\varrho) \quad \quad \text{with}\quad  N_\varrho = M^T \frac{N_\nu}{|d_\nu|^2} M
\end{align}

Lastly, we augment the EHT observations with a concurrent flux measurement of Sgr\,A* from ALMA \parencite{wielgus2022millimeter}.
We achieve this by integrating the flux of our reconstructed source at every point in time, denoted as $s_t = \int_x s$, and comparing it with the observed values. 
We assume a Gaussian likelihood and an uncertainty at the 10\% level, given by:

\begin{align}
\mathcal{P}(d_{\mathrm{lc}}| s) =  \mathcal{N}(d_{\mathrm{lc}}| s_t, N_{\mathrm{lc}})\quad \quad \text{with} \quad N_{\mathrm{lc}} = \widehat{\sigma_{\mathrm{lc}}^2} \quad \text{and}\quad\sigma_{\mathrm{lc}} = 0.1\,|d_{\mathrm{lc}}|
\end{align}
Here the subscript $\mathrm{lc}$ indicates the quantities related to the light-curve from ALMA.

Concretely we use the information from the A1 calibration pipeline for the HI channel from \textcite{wielgus2022millimeter} as additional data in order to constrain the integrated source flux.

The full likelihood is given by the product of the previously derived likelihoods:
\begin{align}
\mathcal{P}(d_\rho, d_\varphi, d_{\mathrm{lc}}| s, g) = \mathcal{N}(d_\rho|\rho + Gg, N_\rho) \:\mathcal{N}(d_\varphi|\varphi, N_\varphi)\: \mathcal{N}(d_{\mathrm{lc}}| s_t, N_{\mathrm{lc}})
\end{align}

In summary, our likelihood model integrates observational data from different sources, accounts for uncertainties and calibration errors, and utilizes the combination of closure phases, self-calibration log-amplitudes, and flux measurements to reconstruct Sgr\,A*.

\paragraph{Model and Prior}\label{sec:prior}
For an effective reconstruction, it is crucial that our signal model is both consistent with the physical properties of the system and flexible enough to capture its complexities. 
The vicinity of black holes is a highly dynamic and complex environment, for which no accurate and expressive parametric models currently exist. 
As such, in this work we employ Gaussian Processes, non-parametric models capable of representing any source configuration.

Nevertheless, we want to impart certain properties to the model, including positivity, spatial and temporal correlation, as well as the ability to reflect exponential changes across linear distances. 
This is achieved by exponentiation of a Gaussian Process random variable $e^\tau$ with $\tau \sim \mathcal{N}(\tau| 0, K)$ with the kernel function $K$ encapsulating the correlation structure in the spatial and temporal domain.
Exponentiation assures positivity as well as exponential behavior.
Another key consideration for this signal model is the interpretability of the model parameters.
An important and intuitive quantity for the signal is the spatially integrated and time-averaged flux of the source, and we want to be able to explicitly express our prior knowledge in it.
This quantity is implicitly modeled by the parameters of the Gaussian Process, but it is distributed over multiple model parameters in highly unintuitive ways.
Therefore, we introduce the time-averaged mean as parameter $b$ with $b\sim \mathcal{LN}(b | \mu_b, \sigma_b^2)$, where $\mathcal{LN}$ indicates a log-normal distribution with respective mean and variance.
To remove model redundancies, we normalize the exponentiated Gaussian Process realization in space and average in time.
Our signal model is therefore given by
\begin{align}
    s = b \frac{e^{\tau}}{\frac{1}{T} \int dt \: dx\: e^{\tau}} \quad \text{.}
\end{align}
This model ensures that we can explicitly state our prior knowledge about the time-averaged flux of the source, while the Gaussian Process allows for flexibility in representing the complex and dynamic structures in the vicinity of the black hole.

For the inference machinery, we utilize Metric Gaussian Variational Inference \parencite{mgvi}. 
This approach relies on a standardized model description, meaning all prior distributions must be expressed in terms of a priori independent, standard normally distributed parameters, $\xi \sim \mathcal{N}(\xi |0, \mathds{1})$. 
This standardization is achieved by reparametrizing the model parameters. 
The general shape of our source, denoted by $\tau$, follows a Gaussian Process. As a (infinite-dimensional) multivariate Gaussian distribution with a specific correlation structure, $\tau$ can be standardized in the same way as in the one-dimensional case via reparametrization:
\begin{align}
    \tau = A \,\xi_\tau
\end{align}
Here, the kernel $K = A \,A^T$ is expressed in terms of a matrix square root (or amplitude matrix) $A$. 
This matrix rotates and scales the input according to the eigenvectors and eigenvalues of the kernel.
We do not need to shift the output, as no prior mean is assumed.

Assuming a priori homogeneity in the spatial and temporal domains of the signal, the correlation kernel $K = \mathds{F}\, \widehat{p} \,\mathds{F}^T$ is, as per the Wiener-Khintchin theorem \parencite{Wiener, Khintchin}, diagonal in the Fourier domain and can be expressed in terms of a power spectrum $p$.
An efficient implementation of a matrix square root for the kernel is therefore $A = \mathds{F} \,\widehat{\sqrt{p}}$.

Given that the spatial and temporal correlation structures stem from physically different processes, we don't expect them to be the same in both dimensions and we assume them to be independent. 
Consequently, we express the amplitude matrix as an outer product of domain-specific amplitude matrices $A = A_t \otimes A_x$.
For more details refer to the method section of \textcite{arras2022variable}.

To address the a priori unknown correlation structure of the individual sub-domains, we incorporate them as a part into the model. 
This correlation structure functions analogously to a regularizer in the Regularized Maximum Likelihood paradigm, and by incorporating its strength into the model, we allow the data to guide the requisite level of regularization. This approach, when coupled with our variational approximation, is intrinsically guarded against over-fitting.

For the modeling of the individual power spectra for both the spatial and temporal domain, we employ descending power-laws and parameterize the slope, the variance in the original domain, as well as several zero-mode variances. 
For more details refer again to the method section of \textcite{arras2022variable}.
Ultimately, all parameters are recast in terms of a priori Gaussian parameters, denoted as $\xi_A \sim \mathcal{N}(\xi_A | 0, \mathds1{})$. 
The slope determines the smoothness of the underlying function, with a power-spectrum slope of \enquote{0} indicating white noise, \enquote{-2} denoting continuous signals, and \enquote{-4} signifying continuously differentiable signals. 
Lower slopes correspond to increasingly smooth signals.

Given the a priori uncertainty regarding the amplitude calibration, we introduce supplementary antenna-based parameters to simulate log-amplitude gains. 
Utilizing the instrument calibration's antenna-based amplitude uncertainties \parencite{ehtiii}, we constrain the log-amplitude gains $g$ within a Gaussian prior.
These gains are subsequently added to the model amplitudes. 
We express them, once again, in terms of standard normal parameters denoted as $\xi_g \sim \mathcal{N}(\xi_g | 0, \mathds{1})$ and $g = \sigma_g \xi_g + \mu_g$.
Our prior choices for all the parameters can be found in \cref{tab:imaging_parameters}.

\paragraph{Variational Inference}\label{sec:mgvi}

The result of the reconstruction is given in terms of the posterior distribution $\mathcal{P}(\xi \vert d)$, which is a probability distribution over tens of millions of parameters.
We do not have access to an analytic solution, and a full exploration with MCMC methods is prohibitively expensive in terms of computational cost.
We instead choose to approximate the true posterior with a more tractable distribution $\mathcal{Q}_\eta(\xi)$. 
The goal is to maximize the information overlap of the true posterior with this approximation. 
We achieve this by minimizing the Kullback-Leibler (KL-) divergence with respect to the parameters of the approximation $\eta$:
\begin{align}
    D_{\mathrm{KL}}\left(\mathcal{Q}_\eta(\xi)||\mathcal{P}(\xi \vert d)\right) = \int d\xi \: \mathcal{Q}_\eta(\xi) \: \mathrm{ln} \left(\frac{\mathcal{Q}_\eta(\xi)}{\mathcal{P}(\xi|d)}\right)
\end{align}
In our case this approximate distribution will be a Gaussian $\mathcal{Q}_\eta(\xi) = \mathcal{N}(\xi | \bar{\xi}, \Xi)$.
As the size of the approximate covariance $\Xi$ scales quadratically with the number of model parameters and we are dealing with millions of those, we will not explicitly parametrize it.
Instead we approximate it with the inverse of the Fisher metric of the posterior evaluated at our current mean estimate $\bar{\xi}$.

In the standardized coordinates we use this covariance has the form $\Xi = {\left( J^T M J + \mathds{1}\right)}^{-1}$.
Here $J$ is the Jacobian of the forward model w.r.t.\ the model parameters, $M$ is the Fisher information metric of the likelihood in the usual coordinates of the distribution and $\mathds{1}$ originates from the Fisher metric of the prior.
We can express this matrix in terms of implicit operations and never have to instantiate it as a dense matrix.
Exploiting the properties of Gaussians and the fact that we optimize the KL-divergence, we can simplify the loss function.
Combining this with an stochastic approximation of the expectation value with $N$ samples, the loss function becomes
\begin{align}
    \underset{\bar{\xi}}{\mathrm{argmin}} \:  D_{\mathrm{KL}}\left(\mathcal{N}(\xi | \bar{\xi}, \Xi)||\mathcal{P}(\xi \vert d)\right) \:\widehat{\approx} \: \: \underset{\bar{\xi}}{\mathrm{argmin}} \: \frac{1}{N} \sum_N - \mathrm{ln}\left(\mathcal{P}(d,\bar{\xi} +\xi_i^*)\right) \\
    \quad \quad \quad \text{with} \quad \xi_i^* \sim \mathcal{N}(\xi | 0, \Xi)\text{.}
\end{align}
This constitutes the Metric Gaussian Variational Inference~(MGVI) approach.
For more details refer to \textcite{mgvi}.
The algorithm iterates between drawing samples $\{\xi_i^*\}_N$ at the location of the current mean estimate $\bar{\xi}$ and optimizing the KL-divergence given this set of samples. 
We iterate the procedure until convergence.
One peculiarity of the approach is the usage of antithetic sampling.
Note that the samples $\xi_i^* \sim \mathcal{N}(\xi | 0, \Xi)$ are zero-centered and an equally valid sample is $-\xi_i^*$, which is totally anti-correlated to the original one.
As we use a significant amount of the computational budget for drawing these samples, we use both of them. 
This has the added benefit that anti-correlation stabilizes expectation values and increases the number of effective samples.

As we approximate the potentially multi-modal posterior distribution with a uni-modal Gaussian, we are not guaranteed to converge towards the global mode. 
In order to counteract this issue and to also partially account for the multi-modality, we repeat the reconstruction several times with different random seeds and starting positions and combine those for the full analysis.

\paragraph{Reconstruction}\label{sec:reconstruction}
The reconstruction procedure itself is rather complex, as we go through several resolutions, likelihoods and optimizers to improve convergence speed and robustness.
Our field of view remains at \SI{200}{\muas} squared and we always consider the full duration of the observation.
Initially we use $64\times 64$~pixels in the spatial domain and a time resolution of about \SI{33}{\min}.
We initialize our model by fitting it to a spatially static disk model with a radius of \SI{45}{\muas}, smoothed edges and a total flux following the ALMA light-curves in time.
This step is important to localize the flux in the center, as the absolute locations are not constrained by the closure phases. 
Without this, we regularly observe random source positions, duplicated sources or honeycomb-like structures throughout the image.
Nevertheless, this issue sometimes persists and we disregard the reconstruction if we observe such behavior in the initial reconstruction steps.
As we do not have an analytic way to initialize the model to this setup, we optimize the model parameters to a noisy version of the disk.

After this initialization step, we start the inference with MGVI using the real likelihoods and 32~sample pairs.
We always include the closure likelihoods, as well as the ALMA light-curve likelihood.
For the first ten iterations of MGVI we start with a mixture of \SI{90}{\percent} closure-log-amplitude- and \SI{10}{\percent} of a self-calibration-log-amplitude-likelihood.
This helps with constraining the overall source morphology while avoiding the over-fitting of the calibration solutions to initially low quality reconstructions.

For the first five MGVI iterations we also use 50~steps of the less aggressive L-BFGS optimizer to also avoid over-fitting, but afterwards we change to 10~steps of the more efficient NewtonCG optimizer.
After the first 10~MGVI iterations, we drop the closure-log-amplitudes and use \SI{100}{\percent} of the self-calibration-log-amplitude likelihood.

After a total of 15~MGVI iterations we consider this low-resolution reconstruction as converged and we increase the resolution of our model.
We do this by interpolating the low resolution result to the next resolution and then fit the higher resolution model to a noisy version of it, analogously to the initialization.
Our next, intermediate, resolution is $96 \times 96$~spatial pixels and a temporal resolution of \SI{5}{\minute}.

Due to the increased computational cost, we reduce the number of sample pairs to~16.
We reset the self-calibration and repeat the same procedure regarding 10~MGVI iterations of \SI{90}{\percent} closure-log-amplitudes, 5~of which with L-BFGS as minimizer and NewtonCG afterwards.
For this intermediate resolution we perform 25~total MGVI iterations.
After this we increase the resolution once again to the final $128 \times 128$~spatial pixels as well as a temporal resolution of \SI{1}{\minute}.
We perform 15~iterations of MGVI with the L-BFGS optimizer and increase the number of optimization steps after the first 5~MGVI iterations from 50 to~100.
We end the reconstruction with 10~more MGVI iterations with the NewtonCG optimizer.\@

Furthermore we gradually decrease the assumed error on the ALMA light-curves from initially \num{50}$\%$ to \num{30}$\%$ and finally \num{10}$\%$ with increasing temporal resolution.
Note that the sampling of this light-curve is higher than our temporal resolution, so we need to inflate the assumed error budget to avoid model inconsistencies. 

Overall we repeat this reconstruction for both days and with 12~different random seeds and initializations.
We perform all subsequent analyses on the aggregate of these reconstructions, which amounts to 384~samples in total for each day.
On April~6th we achieve reduced $\chi^2$ values of $\chi^2_\rho= 1.33$ for the log-amplitudes and  $\chi^2_\varphi=1.50$ for the closure phases.
On April 7th we have $\chi^2_\rho=1.39$ and  $\chi^2_\varphi = 1.86$. 
These values are averaged over all samples and reconstructions and are consistent throughout the reconstructions.
These $\chi^2$ values are expected to be slightly larger than unity, as we do not inflate the assumed error budget to account for unmodelled effects.
Nevertheless, our reconstructed sources accurately explain the data and are in line with values obtained by the EHTC for dynamic reconstructions.





\paragraph{Image Analysis}\label{sec:VIDA}

To extract physical quantities from the reconstructions, we perform Variational Image Domain Analysis \parencite{tiede2022vida} using the VIDA.jl package.
We use it to extract the parameters of Gaussian cosine ring model with added floor for our reconstructed source. 
This model is an extension of the one used by the EHTC for their analysis on Sgr A*, which assumed a circular morphology of the emission. 
Due to the observed non-circular emission in our reconstruction, we relax this assumption and introduce additional ellipticity parameters. 

In the following we briefly summarize the parametric template as outlined in \textcite{tiede2022vida}, following its notation.
The ellipse itself is parametrized in terms of center coordinates $x_0$ and $y_0$. 
The size is described by $d_0$, the geometric mean of the minor and major half axis. 
For a circle this coincides with the diameter.
The ellipticity $\tau$ describes the elongation of the ellipse and $\xi_\tau$ its angle.
Furthermore, its width and azimuthal structure are described by cosine expansions.
\begin{align}
h_{\theta}(x,y) = S_M(\phi(x,y), \boldsymbol{s}, \boldsymbol{\xi}^{(s)})\: \mathrm{exp} \left[\frac{-d_\theta^2(x,y)}{2\sigma_N^2(\phi(x,y),\boldsymbol{\sigma}, \boldsymbol{\xi}^{(\sigma)})}\right] + f(x,y,\gamma,\sigma_1,d_0)
\end{align}
Here $d_\theta(x,y)$ describes the minimum distance of a coordinate to the given ellipse.
Variations along the azimuthal direction are described by
\begin{align}
S_M(\phi(x,y), \boldsymbol{s}, \boldsymbol{\xi}^{(s)}) = 1- \sum_{m=1}^M s_m \:\mathrm{cos}[m(\phi - \xi_m^{(s)})]
\end{align}
in terms of a cosine expansion with amplitude parameters $\boldsymbol{s}$ and phase $\boldsymbol{\xi}^{(s)}$ along the ellipse.
Similarly, 
\begin{align}
\sigma_N(\phi(x,y),\boldsymbol{\sigma}, \boldsymbol{\xi}^{(\sigma)})= \sigma_0 + \sum_{n=1}^N \sigma_n \:\mathrm{cos}[n(\phi - \xi_n^{(\sigma)})]
\end{align}
allows for varying radial width along the azimuthal direction. 
The floor function $f(x,y,\gamma,\sigma_1)$ has a constant value $\gamma$ within the ellipse and tempers off exponentially outside.
\begin{align}
    f(r(x,y),\gamma,\sigma_1,d_0) = \gamma \: \mathds{1}_{\text{in}}(x,y)  + \gamma \:  \mathrm{exp}\left[\frac{-d_\theta^2(x,y)}{2\sigma_1^2}\right] \: (1-\mathds{1}_{\text{in}}(x,y)) 
\end{align}
Here $\mathds{1}_{\text{in}}(x,y)$ indicates whether the coordinates correspond to a location within the ellipse or not.
In order to allow for sufficient flexibility in the morphology, in our analysis we set $N = 10$ and $M = 16$.

As our reconstruction is not just an image, but a dynamically changing object, we perform the parameter extraction for every frame. 
This way we can track changes in the intrinsic source properties in time and characterize its evolution.
Furthermore, the result of our reconstruction is a collection of samples that contain information about the uncertainty of features. 
In order to propagate this uncertainty we have to perform the parameter extraction for all of them individually. 
This then provides an ensemble for each of the model parameters. 
As we also average our results over several independent reconstructions, we have to solve an enormous amount of high-dimensional optimization problems.
This involves one fit for every frame of every sample of every reconstruction and both days.
In order to speed up this procedure, we start by first performing a high-fidelity optimization for a time- and sample-averaged image, which we then use as a starting point for all subsequent samples and frames.
For the initial optimization we run 60~instances of Black Box optimization \parencite{BlackBoxOptim.jl} with \num{25000}~evaluations and select the best-performing parameters.
For the individual frames we only use one instance of the same optimization.
As a further measure to reduce computational costs, we crop the center of the image in order to remove the outer regions without any flux and we re-scale to a resolution of $64 \times 64$~pixels.

\paragraph{Geometric Analysis}\label{sec:hotspot}

The first step in our geometric analysis of the dynamic feature is its extraction from the background, the comparatively static ring emission. 
For this we compute the temporal median of the source on April 6th. 
Any feature that is shorter lived than half of the observational period does not contribute significantly to the median.
We subsequently subtract the temporal median from all frames. We do this for all samples and reconstructions and average the result. 
At this stage we remove any negative values and use this as the target of our geometric model fitting.

We base our geometric model on a simplified version of the one described in \textcite{palumbo2022bayesian} and assume optically thin, axisymmetric and equatorial emission.
The model includes parameters for the orientation of the black hole system with respect to the observer, parameters for the properties of the disk, as well as parameters for the black hole itself.
For simplicity we assume a spectral index of $1$, following \textcite{narayan2021polarized}, which corresponds to a falling flux with frequency.
Here we want to focus on the orientation in terms of inclination and position angle of the projected spin axis, as well as the size and evolution of a localized emission orbiting the black hole.
For the latter we extend the model by assuming a Gaussian shaped emission in radial and a van-Mises shaped emission in the angular direction. 
Both these distributions are parametrized in terms of a location and scale in the respective direction. 
In order to describe the observed orbital motion, we parametrize the location in the angular direction as a linear function in time with some angular velocity and intercept.

We furthermore extend the model in order to account for the observational effects. 
We introduce two shift parameters to correct for translations due to closure quantities and a Gaussian blur parameter to account for limited angular resolution.

Due to a strong anti-correlation between the radial distance of the emission feature and the angular gravitational radius of the black hole, we cannot use this approach to constrain the size and therefore the mass of the black hole.
For a discussion of this anti-correlation, see \textcite{palumbo2022bayesian}. 
Our chosen parameter ranges can be found in \cref{tab:geometric_parameters}.
The assumed range for the gravitational radius $\theta_g$ tightly corresponds to a black hole mass of $4.415 \times 10^6 \mathrm{M}_\odot$ at a distance of $8178 \,\text{pc}$, as constrained by the GRAVITY Collaboration \parencite{gravity2019geometric}.
The remaining parameters are restricted within widely uninformative intervals.

As our analysis requires a large amount of model evaluations, we reduce the spatial resolution to $48 \times 48$ pixels and a temporal resolution to \num{4} minutes in order to significantly reduce required ray-tracing.
For this reason we limit the minimal radial characteristic size to $0.3 \,\text{M}$ in order to ensure that the emission crosses the path of the rays. 
Our model can therefore not represent smaller radial extensions.

Furthermore, we assume a pixel-wise uncertainty with a magnitude of half the highest value in each frame in order to account for extraction and reconstruction artifacts. 
We use this as a likelihood and combine it with the model and parameter priors in order to obtain a posterior distribution over the model parameters given the observed dynamics.
We explore this posterior with Dynamic Nested Sampling \parencite{skilling2006nested, higson2019dynamic} using the dynesty library \parencite{speagle2020dynesty}.
The sampling starts with \num{2000} live points and we initially run it for \num{30 000} iterations. We add \num{200} live points in each batch and run until convergence.
This procedure is repeated \num{50} times with different random seeds to account for the multi-modality of the problem and the individual runs are merged into one overall result.

As we identify two clearly separated modes, we isolate them using a inclination cutoff at $\theta_o  = 115^\circ$.

\begin{table}[h]
\centering

\begin{tabular}{lccc} 
    \toprule
    & April~6th & April 7th & Average \\
    \midrule
    Radius $r$ $(\mu\text{as})$ & $23.94 \pm 0.67$ & $25.57 \pm 0.74$ & $25.00 \pm 1.05$ \\
    Width $w$ $(\mu\text{as})$ & $12.03 \pm 0.49$ & $12.56 \pm 0.46$ & $12.38 \pm 0.54$ \\
    Asymmetry $A$ & $0.34 \pm 0.14$ & $0.16 \pm 0.09$ & $0.22 \pm 0.14$ \\
    Asymmetry Angle $\eta$ $(^\circ)$ & $198.77 \pm 32.66$ & $335.63 \pm 64.42$ & $325.20 \pm 98.56$ \\
    Floor Ratio $f_c$ & $0.34 \pm 0.16$ & $0.13 \pm 0.11$ & $0.21 \pm 0.17$ \\
    Ellipticity $\tau$  & $0.37 \pm 0.04$ & $0.32 \pm 0.04$ & $0.34 \pm 0.05$ \\
    Ellipticity Angle $T$ $(^\circ)$ & $7.38 \pm 10.47$ & $10.91 \pm 10.24$ & $9.68 \pm 10.73$ \\
    \bottomrule
\end{tabular}
\caption[Image analysis results.]{The results of the image domain analysis with VIDA for the two individual days as well as their average.
    The uncertainties are provided as one standard deviation. These values are time- and sample-averaged and the indicated errors therefore contain both sources of variance.}
\label{tab:simple_table}
\end{table}

\begin{table}
    \centering
    \begin{tabular}{lc} 
    \toprule 
    Hyperparameter & Prior or Value \\
    \midrule 
    Number of pixels (along edge) & $N_{\mathrm{pix}} = 128$ \\
    Field of view ($\mu\text{as}$ along edge) & $\mathrm{FOV} = 200$ \\
    Temporal resolution (minutes) & $t_r=1$ \\
    & \\
    Zero mode variance& $\alpha \sim \mathcal{LN}(\alpha| 1,0.3)$  \\
    & \\
    Spatial fluctuations & $a^{(x)}\sim \mathcal{LN}(a^{(x)}| 1.5,1)$ \\
    Spatial correlation power-law &$m^{(x)}\sim \mathcal{N}(m^{(x)}| -3,1)$  \\
    &  \\
    Temporal fluctuations& $a^{(t)}\sim \mathcal{LN}(a^{(t)}| 0.5,0.2)$ \\
    Temporal correlation power-law& $m^{(t)}\sim \mathcal{N}(m^{(t)}| -2,0.5)$ \\
    &  \\
    Time-averaged total flux& $b \sim \mathcal{LN}(b| 2.5,1)$\\
    log-Amplitude gain variances& $g \sim \mathcal{N}(g|0,\sigma_g^2)$ with $\sigma_g$ from table~3 in \parencite{ehtiii}.\\

    \bottomrule 
    \end{tabular}
    \caption[Imaging model parameters.]{The parameter priors and hyperparameter values for our imaging model. 
    Quantities without distributions are fixed, Here $\mathcal{LN}$ indicates a log-normal distribution with respective mean and variance.
    For the role of the zero mode variance, power-law, and fluctuation parameters refer to the methods section of \textcite{arras2022variable}.}
    \label{tab:imaging_parameters}
\end{table}

\begin{table}
    \centering
    \begin{tabular}{lcccc} 
        \toprule 
        Ray Tracing Parameters & Symbol & Prior Range & Edge-On Mode& Face-On Mode\\
        \midrule 
        Angular gravitational radius ($\mu\text{as}$) & $\theta_g$ & $[4.95, 5.05]$ & ${4.98}^{+0.03}_{-0.03}$  & ${4.99}^{+0.06}_{-0.04}$ \\
        Observer inclination ($^\circ$) & $\theta_o$ & $[0, 180]$ &${79.6}^{+1.2}_{-6.9}$ &${159.4}^{+14.3}_{-8.1}$ \\
        Dimensionless spin & $a_*$ & $[-1, 1]$ & ${-0.81}^{+0.10}_{-0.03}$& ${0.25}^{+0.19}_{-0.62}$\\
        Field of view ($\mu\text{as}$ along edge) & FOV & 160 & & \\
        Number of pixels (along edge) & $N_{\mathrm{pix}}$ & 48 & & \\
        Time duration (minutes) & $T$ & $100$ \\
        Temporal resolution (minutes) & $t_r$ & 4 & & \\
        Maximum photon winding number  & $n_\mathrm{max}$ & 1 & & \\
         & & \\
        Characteristic radius ($M$) & $r_c$ & $[1.5,15]$ & ${4.70}^{+0.59}_{-0.24}$ & ${4.67}^{+0.82}_{-0.33}$\\
        Characteristic radial size ($M$) & $\sigma_r$ & $[0.3,2]$ & ${0.51}^{+0.31}_{-0.19}$&${1.41}^{+0.28}_{-1.11}$ \\
        Characteristic angle ($^\circ$) & $\phi_c$ & $ \phi_t t + \phi_0$  & &\\

        Characteristic angular size ($^\circ$) & $\sigma_\phi$ & $[0.063,360]$  & ${15.5}^{+3.5}_{-2.9}$& ${11.0}^{+2.0}_{-2.7}$\\
        Angular velocity ($\frac{^\circ}{\mathrm{min}}$) & $\phi_t$ & $[-6.3,6.3]$  & ${-2.12}^{+0.23}_{-0.29}$& ${1.93}^{+0.28}_{-0.32}$\\
        Angular intercept ($^\circ$) & $\phi_0$ & $[-360,360]$  &${63}^{+75}_{-50}$ & ${309}^{+45}_{-298}$\\
         & & \\
        Fluid speed (fraction of c) & $\beta$ & $[0,0.9]$  & ${0.24}^{+0.15}_{-0.15}$& ${0.81}^{+0.09}_{-0.48}$\\
        Equatorial fluid velocity angle ($^\circ$)& $\chi$ & $[0,90]$ & ${115}^{+32}_{-86}$ & ${116}^{+32}_{-86}$\\
        Vertical magnetic field angle ($^\circ$) & $\iota$ & $[-180,180]$ & ${1}^{+169}_{-171}$ & ${1}^{+168}_{-171}$\\
        Spectral index & $\alpha_\nu$ & $1$ & &\\
         & & \\
        Total Stokes I flux & $I_\mathrm{tot}$ & $1$  & &\\
        Projected spin axis position angle ($^\circ$) & PA & $[-90, 90]$& ${0.4}^{+1.6}_{-3.0}$ & ${-19.5}^{+2.9}_{-3.2}$   \\
         & & \\
        Shift x  ($\mu\text{as}$)& $\Delta_x$ & $[-30,30]$ &${-13.7}^{+2.3}_{-0.7}$& ${6.7}^{+2.6}_{-1.4}$ \\
        Shift y  ($\mu\text{as}$)& $\Delta_y$ & $[-30,30]$  & ${2.3}^{+0.9}_{-1.7}$& ${0.5}^{+0.9}_{-0.8}$\\
        Gaussian blur  ($\mu\text{as}$)& $b$ & $[3.5,15]$  &${10.8}^{+0.3}_{-0.4}$ & ${9.1}^{+1.7}_{-0.8}$\\
        \bottomrule 
    \end{tabular}
    \caption[Geometric model parameters and results.]{The parameters and parameter ranges of our geometric model. 
    The ranges indicate a uniform interval within which is optimized. 
    The absence of a range indicate fixed values. 
    The parameter ranges for the posterior modes correspond to 68th~percentiles.
    The layout of this table is based on table~1 in \textcite{palumbo2022bayesian}.}
    \label{tab:geometric_parameters}
\end{table}

\section*{Data Availability}
The data this work is based on have been published by the Event Horizon Collaboration and are available at~\cite{sgradata}.
The reconstructions presented in this work are made available at~\cite{samples}.
\section*{Code availability}
The software sources used for producing the results of this publication are available at~\cite{zenodo_software}.

\sloppy
\printbibliography{}
\fussy

\section*{Acknowledgements}
We thank Maciek Wielgus and collaborators for graciously providing the ALMA light-curves in order to constrain the total flux in our reconstructions.
We furthermore thank Freek Roelofs for organizing ngEHT Analysis Challenges, which provided the perfect environment for the development of our dynamic reconstruction approach.
We thank Oliver Schulz and the Max Planck Institute for Physics for providing computational support and resources.
We thank Martin Reinecke for helpful discussions and comments on the manuscript.
The authors acknowledge the use of the GPT-4 language model \parencite{gpt4}, which assisted in the drafting and editing of this manuscript as well as with smaller coding exercises.
J.K.\ acknowledges the financial support by the Excellence Cluster ORIGINS, which is funded by the Deutsche Forschungsgemeinschaft (DFG, German Research Foundation) under Germany's Excellence Strategy -- EXC-2094-390783311.
P.A.\ acknowledges the financial support by the German Federal Ministry of Education and Research (BMBF) under grant 05A17PB1 (Verbundprojekt D-MeerKAT). %

\section*{Author Contributions}

\section*{Competing interests}
The authors declare no competing interests.

\end{document}